\pdfoutput=1
\documentclass[review]{elsarticle}

\usepackage{lineno,hyperref}
\usepackage{amsmath}
\usepackage{graphicx}
\usepackage{caption}
\usepackage{subcaption}
\usepackage[section]{placeins}

\journal{arxiv.org}

\begin{document}

\begin{frontmatter}

\title{COVID-19 Economic Policy Effects on Consumer Spending and Foot Traffic in the U.S.}

\author[a,b]{Zhiqing Yang}

\author[c]{Youngjun Choe\corref{mycorrespondingauthor}}
\cortext[mycorrespondingauthor]{Corresponding author}
\ead{ychoe@uw.edu}

\author[c]{Matthew Martell}

\address[a]{Department of Statistics, University of Washington}

\address[b]{Information School, University of Washington}

\address[c]{Department of Industrial \& Systems Engineering, University of Washington}

\begin{abstract}
To battle with economic challenges during the COVID-19 pandemic, the US government implemented various measures to mitigate economic loss. From issuance of stimulus checks to reopening businesses, consumers had to constantly alter their behavior in response to government policies. Using anonymized card transactions and mobile device-based location tracking data, we analyze the factors that contribute to these behavior changes, focusing on stimulus check issuance and state-wide reopening. Our finding suggests that stimulus payment has a significant immediate effect of boosting spending, but it typically does not reverse a downward trend. State-wide reopening had a small effect on spending. Foot traffic increased gradually after stimulus check issuance, but only increased slightly after reopening, which also coincided or preceded several policy changes and confounding events (e.g., protests) in the US. We also find differences in the reaction to these policies in different regions in the US. Our results may be used to inform future economic recovery policies and their potential consumer response.
\end{abstract}

\begin{keyword}
\texttt{COVID-19}\sep economic impact \sep interrupted time series analysis
\end{keyword}

\end{frontmatter}


\section*{Introduction}

COVID-19 resulted in a global pandemic. It not only presented a public health crisis, but also brought unprecedented economic challenges. In the US, the unemployment rate shot up from 3.8\% in February 2020 to 13.0\% in May 2020 \cite{kochhar}. To decrease the rate of new cases, US state governments implemented lockdown policies during the initial stages of the pandemic. Consumers began to change their spending patterns to respond to these policies and keep themselves safe, especially in sectors such as restaurants and retail. Foot traffic declined 60\% over March to April, 2020 \cite{Cronin2020}. Self-imposed precautionary behavior accounts for a large portion of the overall drop in foot traffic, suggesting that stay-home orders served as an informational treatment to warn citizens the severity of the virus locally, in addition to formally restricting mobility \cite{Cronin2020}.

After the initial lockdown, the federal and state governments implemented various measures to mitigate economic loss, including stimulus packages and later limited reopening of places of business. Various working papers and publications have analyzed the effects of government policies on consumer behavior. Coibion et al. \cite{coibion} reported a survey study investigating the effect of the timing of \textit{lockdown} policies on households’ spending and macroeconomic expectations. Baker et al. \cite{baker_stimuluscheck} and Chetty et al. \cite{Chetty} found that spending increased sharply following the receipt of a stimulus check, especially among low-income households. Complementing this finding, another survey study \cite{coibion_2} found that only 15 percent of people reported that they mostly spent the money from the stimulus, where others saved or paid off debts. Rahman et al. also explored socioeconomic factors that are associated with sentiments of people about reopening the economy using Twitter and Census data, and found that people with lower education level, lower income, and higher house rent are more interested in reopening the economy \cite{Rahman2020}. These studies were generally focused on the relatively short-term effects of lockdown or stimulus policies with respect to certain consumer socio-demographics in limited geographies. 

Other studies have focused specifically on mobility during the pandemic. Location-based data were widely used in these studies to understand human mobility, including cell phone and GPS data. Beria et al. \cite{Beria2021} examined movements and territorial differences during the lockdown in Italy. Fu et al. \cite{Fu2021} analyzed the spatial and temporal variations in stay-at-home behaviors against social vulnerability indicators, and highlighted the importance of understanding spatiotemporal pattern of public health behaviors in policy-making. Hu et al. \cite{Hu2021} developed a big-data-driven analytical framework and showed that lockdown policies and the reopening guidelines play limited roles in affecting human mobility. They calculated mobility change based on the week after the first reopening guideline was issued by the federal government (April 20th - April 27th, 2020). However, because each state had different reopening dates and policies, it is important to consider state-specific reopening dates as in this study. Furthermore, existing mobility studies mainly focused on lockdown or stay-at-home orders. Few studies examined the effects of later policies for economic recovery. 

While several studies have analyzed the mobility pattern and spending pattern at the beginning of the pandemic, there remain gaps in understanding and comparing the foot traffic and spending changes after fiscal stimulus and reopening policies during the economic crisis. Additionally, different geographical regions warrant comparative analyses because their demographics, socioeconomic status, and political interests vary. In this study, we analyze the consumer response to government policies in mitigating economic loss across US. Using data from SafeGraph, we perform interrupted time series analysis (ITSA) to analyze the effect of these policies on spending and foot traffic. We facilitate our discussion around national trends by examining one state from each of four regions in the US, then expand the analysis on all fifty states. We hypothesize that stimulus checks lead to a sharp increase in spending and that state reopening leads to increased foot traffic. We aim to quantify the immediate (within a few days) and longer-term (within a few months) effects of government measures on two kinds of responses, consumer spending and foot traffic. 

Our result builds on the emerging literature in several aspects. First, we expand the analysis of economic recovery and reopening strategies in the COVID-19 pandemic \cite{alex,Chang,Hasselwander2021,Donoghue}. Our finding stands with Alexander \cite{alex} that there is a relationship between the timing of statewide reopening and consumer responses. Second, we add to the body of literature on examining the effect of expansionary fiscal policies on economic activity \cite{Casado2020,Reichling,Lumpur2009}. We find that providing financial support is a great way of boosting spending in the short run, but the impact is temporary. Last but not least, we build on the analysis of policy interventions during the COVID-19 pandemic, including the effects on energy returns \cite{Saif-Alyousfi2021}, stock market \cite{Narayan}, and employment \cite{Chetty}. We mainly focus on analyzing consumer behavior changes responding to government policies and their regional differences.

\section*{Data and Methods}
We obtained foot traffic data from SafeGraph,\footnote{SafeGraph website: https://www.safegraph.com} a company that provides foot traffic data from opt-in smartphone GPS tracking, which has information on daily visits of 6 million points of interest across the country. We quantify foot traffic by aggregating daily visits in each state in the US from February 2020 to August 2020. The card transactions data are from Facteus,\footnote{Facteus website: https://www.facteus.com. Note that these transactions are made from challenger banks, payroll cards, and government cards, which are mainly used by lower-income and younger populations.} which has been synthesized to protect individual privacy while retaining 99.97\% of the statistical attributes of the original data.  We measured total spending by aggregating all of the transactions per day. 

We first selected one state from the main regions in the US, including Washington from the West Coast, New York from the East Coast, Iowa from the Midwest, and Texas in the South, to conduct our analysis. We then analyzed all 50 states and presented the results visually in a US map. We used April 15th as the date when people started to receive stimulus checks. Since each state had their own reopening policy, we obtained statewide reopening dates from COVID-19 US State Policies (CUSP) database.\footnote{CUSP: https://statepolicies.com}. These reopening dates range from April 20th to June 8th. 

To identify factors that contributed to changes in consumer spending pattern, we conducted an ITSA on the date of issuing stimulus checks nationally, and the first state-ordered reopening date. Before conducting ITSA, we performed seasonality adjustment to remove weekly cycles from the data. We then used a \textit{Segmented Regression Analysis}, which is one of the statistical models for analyzing interrupted time series \cite{Wagner}. We used \textit{intervention1} to represent the issuance of stimulus checks, and \textit{intervention2} to represent the state-ordered reopening. The trend within each of the three segments is approximated by a linear regression line. The model is as follows: 

\begin{equation} \label{eq:1}
\begin{aligned} 
Y_{t}=& \beta_{0}+\beta_{1} \times \text {time}_{t}+\beta_{2} \times \text {intervention} 1_{t} \\
&+\beta_{3} \times \text {time after intervention} 1_{t} \\
&+\beta_{4} \times \text {intervention} 2_{t} \\
&+\beta_{5} \times \text {time after intervention} 2_{t}+\epsilon_{t},
\end{aligned} 
\end{equation}

where $Y_t$ is the measure of total spending in day $t$; $\text{time}_{t}$ is a continuous variable indicating the number of days in \textit{t} since the start of observation which is March 2, 2020; $\text{intervention} 1_{t}$ is an indicator variable coded 0 before April 15, 2020 and coded 1 on and after it; $\text{intervention} 2_{t}$ is an indicator variable coded 0 before the date of reopening (differs across different states) and coded 1 on and after it.
$\beta_{0}$ is the pre-intervention baseline level, and $\beta_{1}$ is the slope of daily spending before the first intervention. $\beta_{2}$ estimates the level change, or immediate effect of the first intervention, and $\beta_{3}$ estimates the change in the slope of daily spending. Thus, $\beta_1 + \beta_3$ is the slope of the period after intervention1, i.e., the average daily increase or decrease of spending in this period. Similarly, $\beta_4$ and $\beta_5$ represent the level change and slope change due to the second intervention, respectively. $\epsilon_t$ represents the variability not explained by the model.

For foot traffic data, we added an additional intervention, which is the closure of nonessential businesses. We denote it by $\text{intervention} 1_{t}$. The stimulus check issuance and reopening of businesses will be $\text{intervention} 2_{t}$ and $\text{intervention} 3_{t}$, respectively. The model is similar to equation \ref{eq:1}, with extra $\beta_6$ and $\beta_7$ that represent level change and slope change, respectively, due to the third intervention. It is shown as follows:

\begin{equation} \label{eq:2}
\begin{aligned}
Y_{t}=& \beta_{0}+\beta_{1} \times \text {time}_{t}+\beta_{2} \times \text {intervention} 1_{t} \\
&+\beta_{3} \times \text {time after intervention} 1_{t} \\
&+\beta_{4} \times \text {intervention} 2_{t} \\
&+\beta_{5} \times \text {time after intervention} 2_{t} \\
&+ \beta_{6} \times \text {intervention} 3_{t} \\
&+ \beta_{7} \times \text {time after intervention} 3_{t}
+\epsilon_{t},
\end{aligned}
\end{equation}

The equations to calculate the percent of level change and slope change for spending and foot traffic are shown in the Appendix.

\section*{Results}
\begin{figure}[h]
\centering
\includegraphics[width=1.0\linewidth]{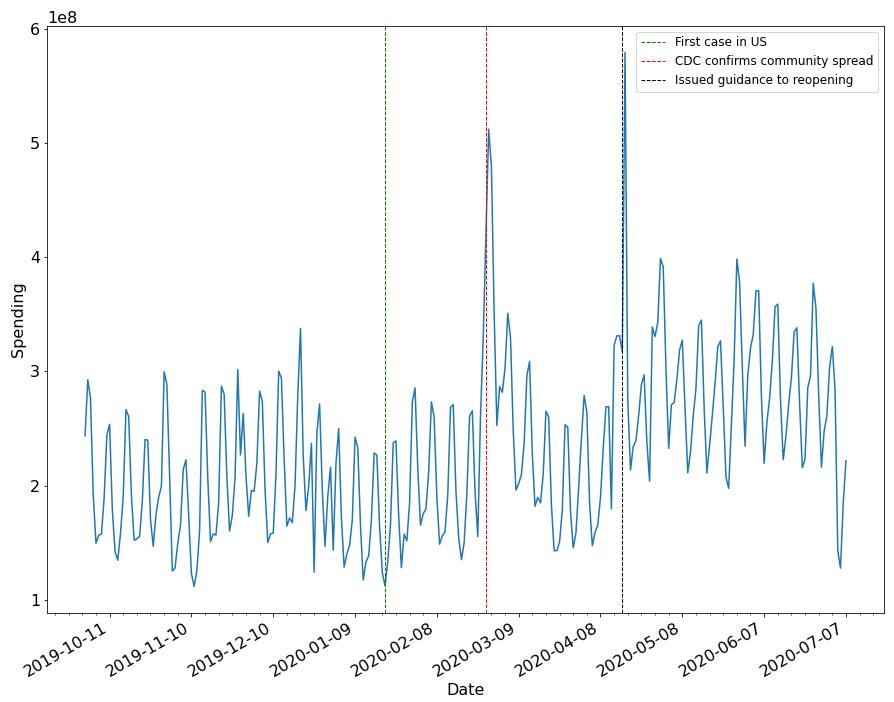}
\caption{Card spending (in USD) from September 2019 to August 2020 in the US. The green vertical dashed line indicates January 20th, when the first COVID-19 case was reported in the US \cite{Holshue}. The red vertical dashed line indicates February 26th, when CDC confirms possible instance of community spread in the US \cite{cdc}. The black vertical dashed line indicates April 17th, when the White House issued reopening guidance to states \cite{Subramanian}.}
\label{fig:spending pattern overall}
\end{figure}

Aggregating total spending in the US, we found that the card spending during the pandemic increased sharply after February 26th, then it dropped back to the baseline level (Fig.~\ref{fig:spending pattern overall}). There are two peaks in the card transactions data: February 26th was the day when many people started panic buying, and April 13th was the day when the first round of stimulus checks was deposited.

\begin{figure}[ht!]
\centering
\includegraphics[height=0.4\textheight]{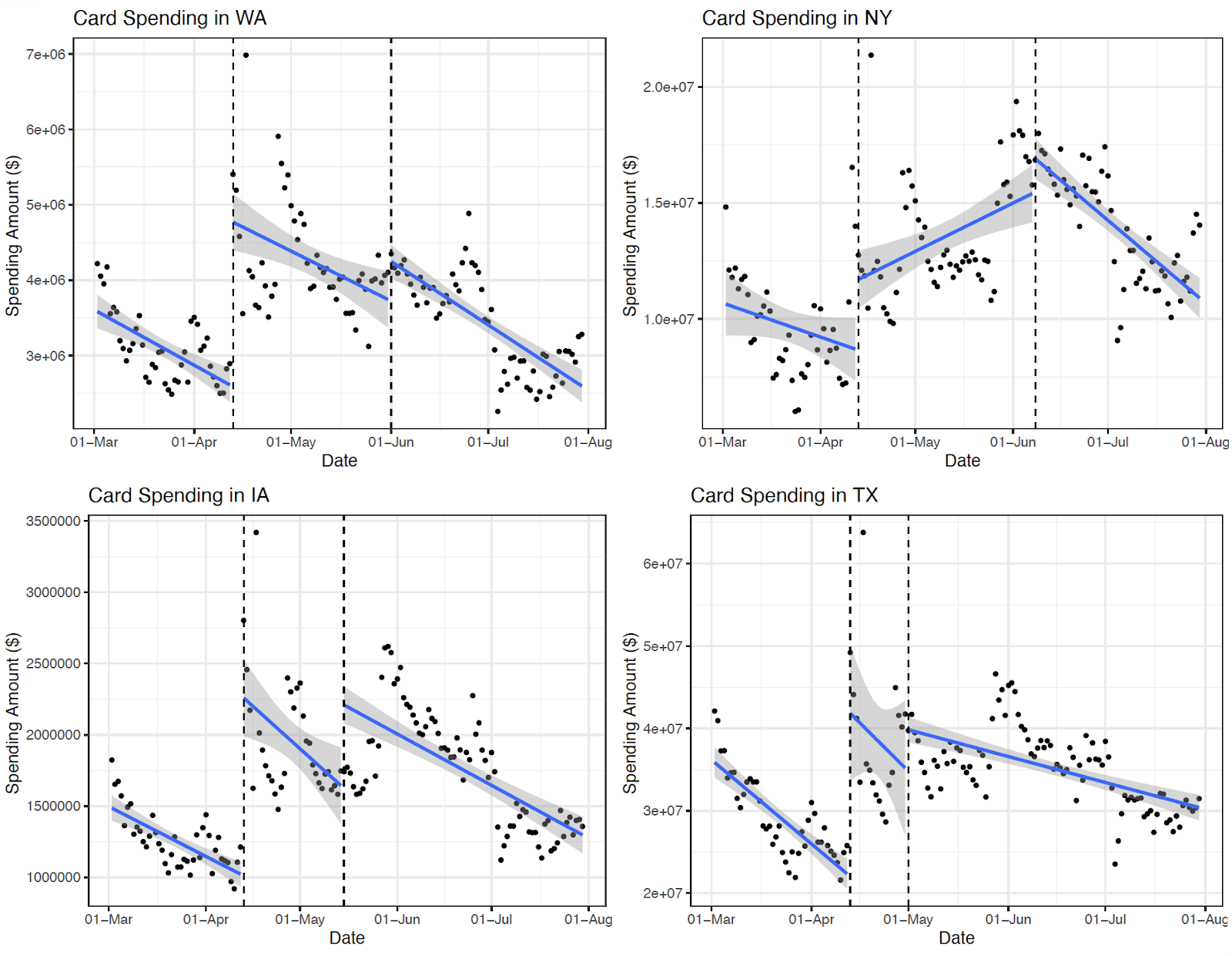}
\caption{ITSA of card spending data, including Washington from the West Coast, New York from the East Coast, Iowa from the Midwest, and Texas in the South. The first vertical dashed line indicates the date of stimulus check issuance. The second vertical dashed line indicates the date of state-wide reopening.}\label{fig:spending_by_state}
\includegraphics[height=0.4\textheight]{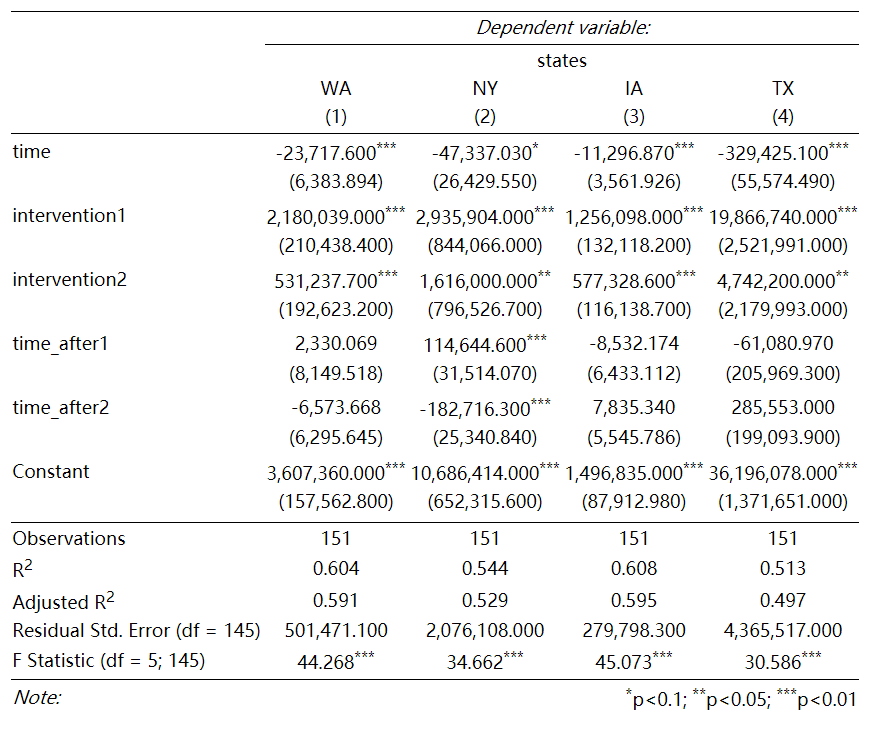}
\caption{Regression table for spending data, using equation \ref{eq:1}. "intervention1" represents issuing of stimulus check, and "intervention2" represents reopening date for each state. }\label{tab:spending_by_state}
\end{figure}

From our ITSA results of spending in the four states (Fig.~\ref{fig:spending_by_state} and \ref{tab:spending_by_state}), the estimated coefficients of \textit{intervention1} are all large, with \textit{p}-values less than $2\mathrm{e}{-16}$ (for New York, the \textit{p}-value is $6\mathrm{e}{-4}$), meaning there is an immediate significant jump in consumer spending after receiving the stimulus checks. For New York state, the estimated slope of spending shifts from negative to positive (\textit{p}-value is $4\mathrm{e}{-4}$), but the slope changes in the three other states are not significant. Although the reopening date varies by state, the immediate effects of \textit{intervention2} are all significant among the four states, but the amount of change is much less than that of \textit{intervention1}. New York and Texas have \textit{p}-value equal to approximately 0.04; Washington has \textit{p}-value equal to 0.006, and Iowa has \textit{p}-value of $1.85\mathrm{e}{-6}$. The change of slope after the second intervention is not significant in any state except for New York, where the spending slope decreases significantly.

For foot traffic data ((Fig.~\ref{fig:spending_by_state} and \ref{tab:spending_by_state})), we added an extra intervention because the trend before the issuance of stimulus checks is not consistent in three of the states -- there is an upward trend before the closure of nonessential businesses, then it immediately sinks (Washington, New York, and Iowa all have a positive $\beta_1$). The closure of nonessential businesses has an immediate significant decrease in foot traffic in New York and Texas. The negative change in slope is also significant, except for Texas, because Texas closed nonessential businesses at a relatively later date, and foot traffic was already decreasing prior to the intervention ($\beta_1$ is negative with a \textit{p}-value \textless 0.01). The immediate positive effect of the issuance of stimulus checks is only significant for Washington and Iowa (\textit{p}-value \textless 0.1). The slope change is significant in all four states, except for Texas. The immediate effect of  \textit{intervention3} is significant in the four states, but the increase gradually flattens, with a negative change in the slope.

\begin{figure}[ht!]
\centering
\includegraphics[height=0.4\textheight]{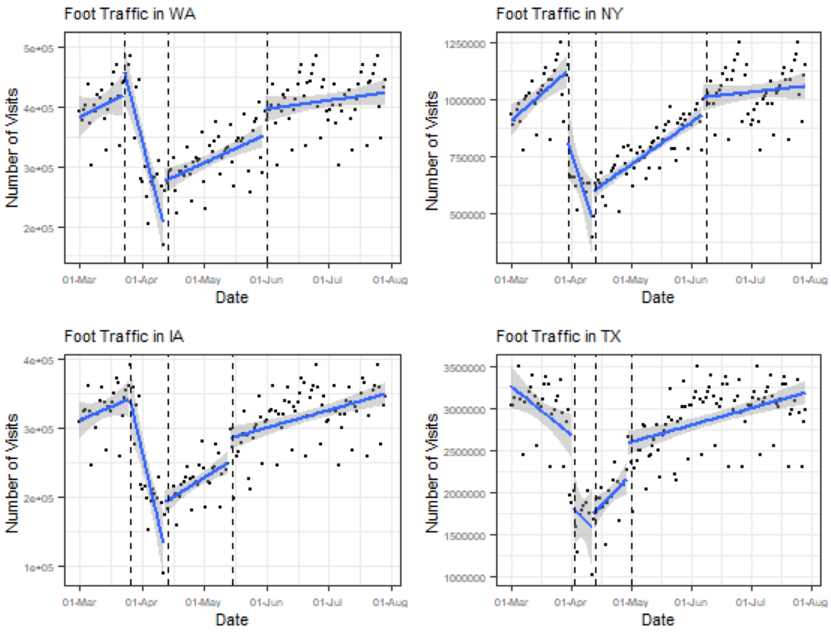}
\caption{ITSA of foot traffic data. The first vertical dashed line indicates the dates when states closed nonessential businesses. The second vertical dashed line indicates the date of stimulus check issuance. The third vertical dashed line indicates the date of state-wide reopening.}\label{fig:traffic_by_state}
\includegraphics[height=0.4\textheight]{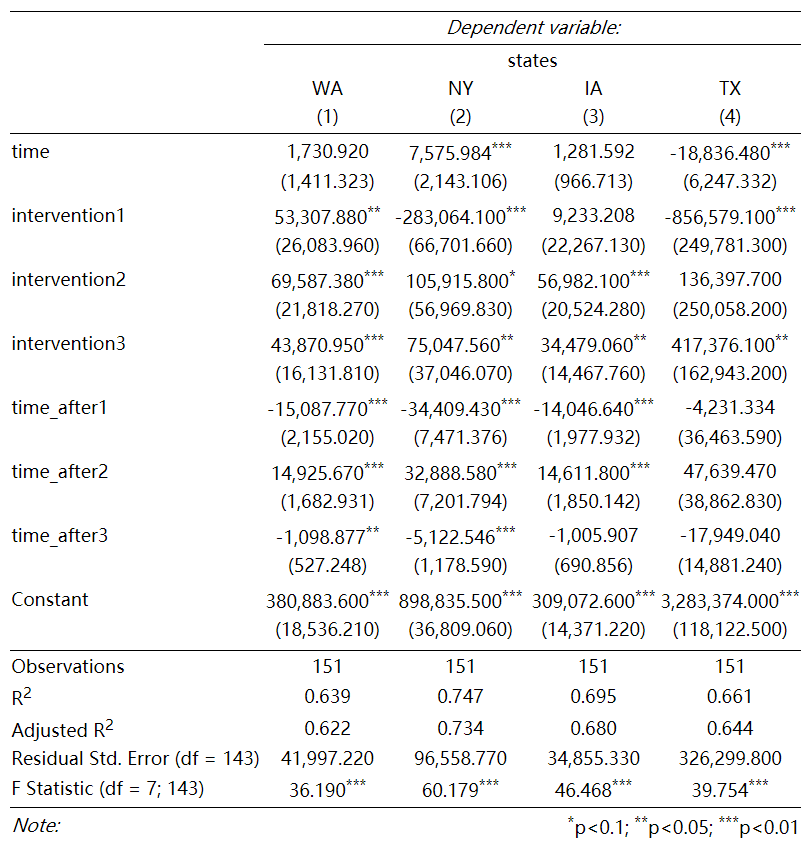}
\caption{Regression table for foot traffic data, using equation \ref{eq:2}. "Intervention1" represents closure of nonessential businesses, "intervention2" represents issuing of stimulus check, and "intervention3" represents reopening date for each state. }\label{tab:traffic_by_state}
\end{figure}

\begin{figure}[ht!]
\makebox[\linewidth][c]{%
\begin{subfigure}[b]{.7\textwidth}
\centering
\includegraphics[width=1\textwidth]{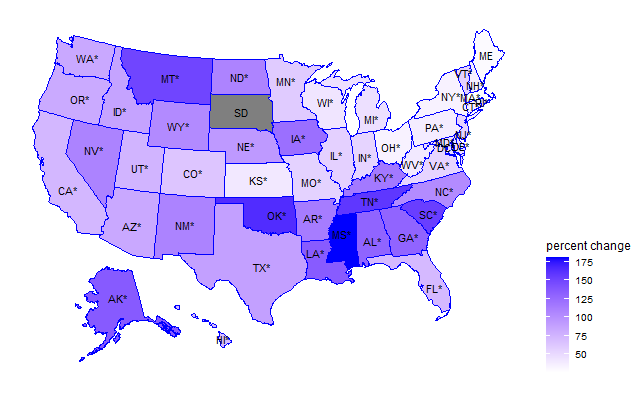}
\caption{First intervention (stimulus checks) level change}
\end{subfigure}%
\begin{subfigure}[b]{.7\textwidth}
\centering
\includegraphics[width=1\textwidth]{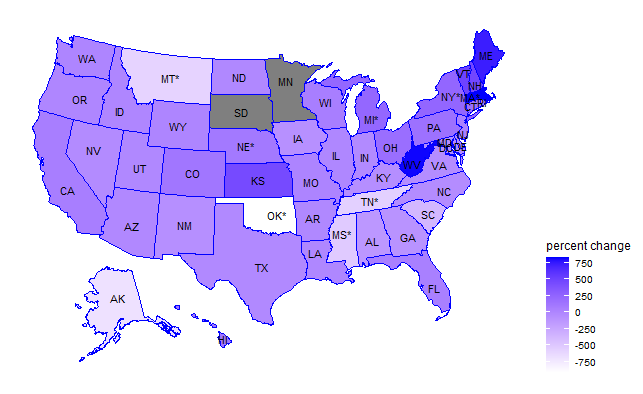}
\caption{First intervention (stimulus checks) slope change}
\end{subfigure}%
}\\
\makebox[\linewidth][c]{%
\begin{subfigure}[b]{.7\textwidth}
\centering
\includegraphics[width=1\textwidth]{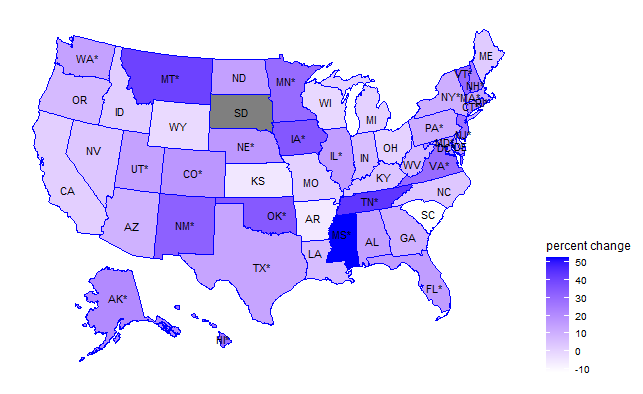}
\caption{Second intervention (reopening) level change}
\end{subfigure}%
\begin{subfigure}[b]{.7\textwidth}
\centering
\includegraphics[width=1\textwidth]{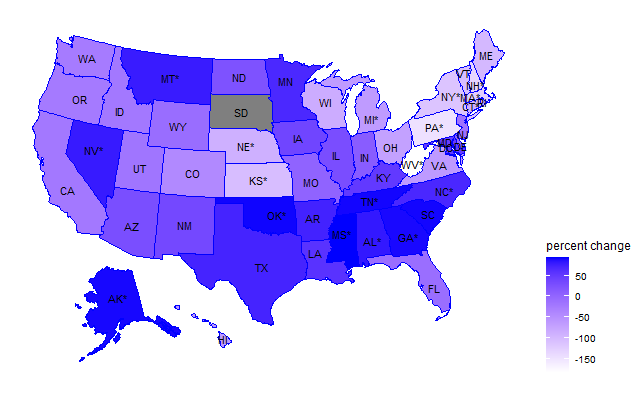}
\caption{Second intervention (reopening) slope change}
\end{subfigure}%
}
\caption{Effects of interventions on spending. The percent changes are represented by the color scheme on the bottom right side of each map. The asterisk next to the state abbreviation indicates that the change is statistically significant ($p < 0.05$). We did not find social distancing or reopening date for South Dakota, so it's colored grey.}
\label{tab:all_spending_intervention}
\end{figure}

\begin{figure}[h!]
\makebox[\linewidth][c]{%
\begin{subfigure}[b]{.7\textwidth}
\centering
\includegraphics[width=1\textwidth]{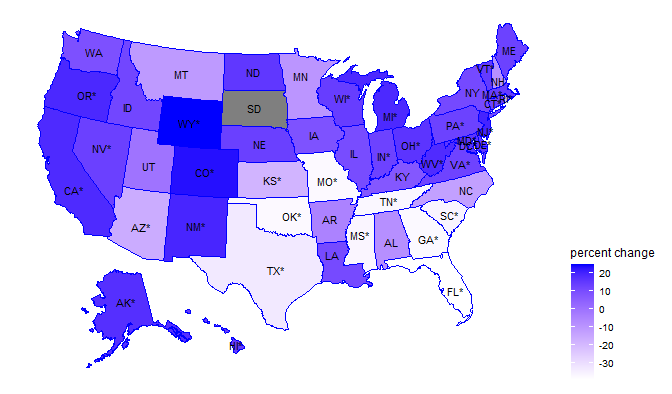}
\caption{First intervention (lockdown) level change}
\end{subfigure}%
\begin{subfigure}[b]{.7\textwidth}
\centering
\includegraphics[width=1\textwidth]{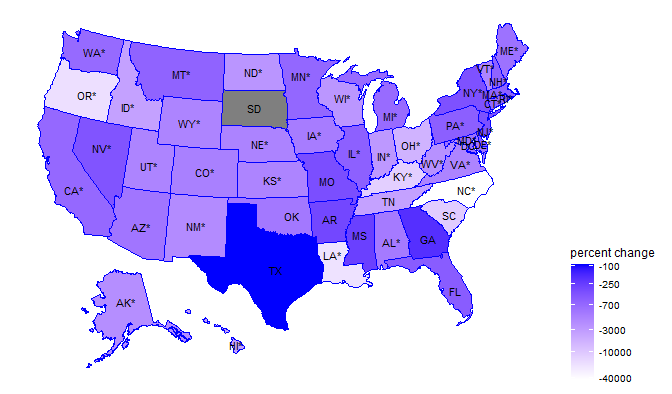}
\caption{First intervention (lockdown) slope change}
\end{subfigure}%
}\\
\makebox[\linewidth][c]{%
\begin{subfigure}[b]{.7\textwidth}
\centering
\includegraphics[width=1\textwidth]{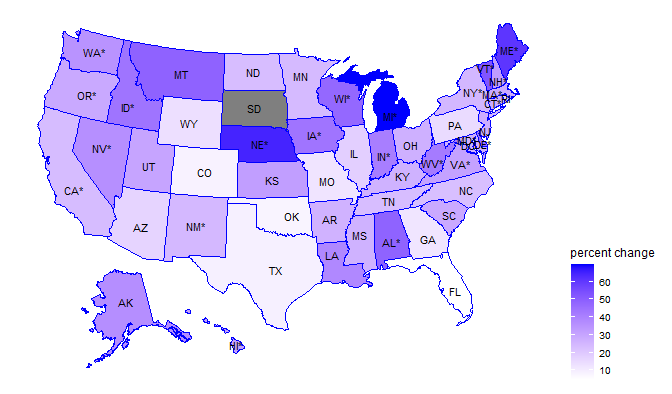}
\caption{Second intervention (stimulus checks) level change}
\end{subfigure}%
\begin{subfigure}[b]{.7\textwidth}
\centering
\includegraphics[width=1\textwidth]{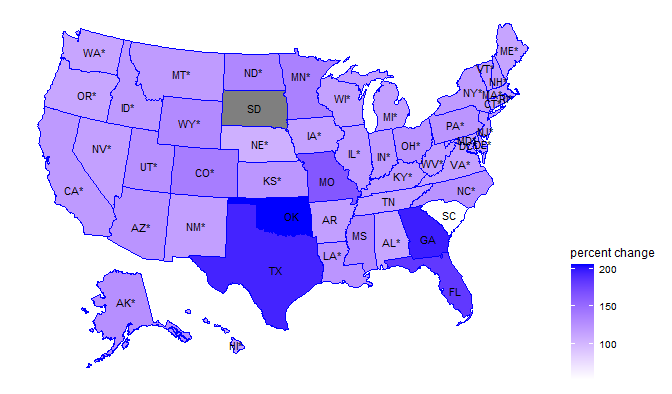}
\caption{Second intervention (stimulus checks) slope change}
\end{subfigure}%
}\\
\makebox[\linewidth][c]{%
\begin{subfigure}[b]{.7\textwidth}
\centering
\includegraphics[width=1\textwidth]{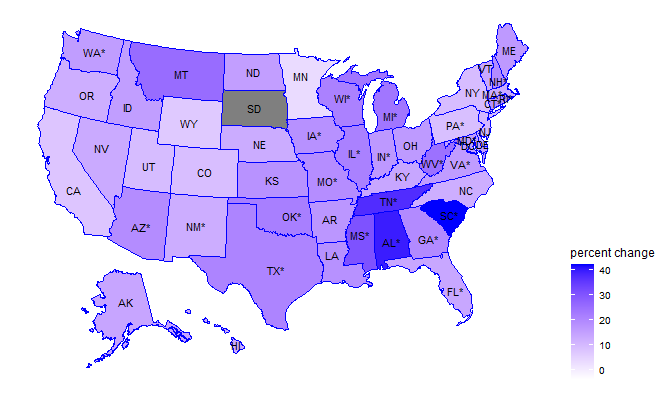}
\caption{Third intervention (reopening) level change}
\end{subfigure}%
\begin{subfigure}[b]{.7\textwidth}
\centering
\includegraphics[width=1\textwidth]{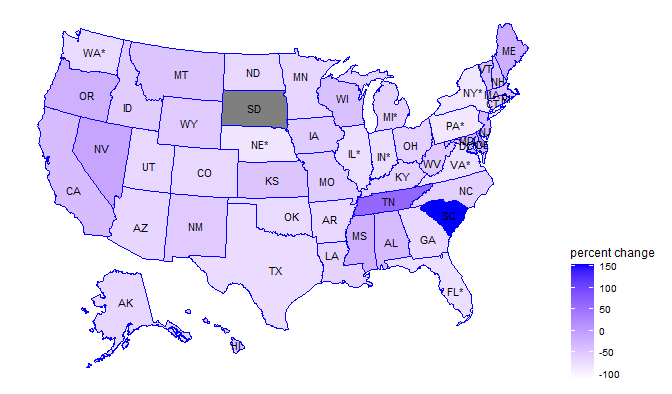}
\caption{Third intervention (reopening) slope change}
\end{subfigure}%
}
\caption{Effects of interventions on foot traffic. See the caption of Fig.~\ref{tab:all_spending_intervention} for interpretation.}
\label{tab:traffic_intervention_all}
\end{figure}

We show the highlights of the results from the rest of the country in Fig. \ref{tab:all_spending_intervention} and Fig. \ref{tab:traffic_intervention_all} . All states experienced an immediate significant increase in consumer spending after receiving the stimulus checks, with an increase of over 50\%. The states in the southeast region had a much greater increase of over 100\%. The percent slope change differs by state. Although it ranges from -750\% to 750\%, many of them are not significant. States like Oklahoma and Mississippi had a significant decrease of -910\% and -518\%, respectively. Minnesota also had a large negative slope change of -21822\%, so we do not include it in the map. The percent increase in spending after the second intervention was much less for most states, ranging from 0 to 50\%, and only half of these changes were significant. The southeastern region typically had a significant increase of over 50\%, but the changes in other states are not significant.

For foot traffic, the effect of lockdown has an immediate positive effect on states on the west coast and northeast region, presumably due to stockpiling of groceries in response to the pandemic. Then the slope decreased sharply for all of the states, especially in North Carolina, which has a -42310\% change in the slope. After issuing the stimulus checks, we see a slight immediate increase in foot traffic, and this effect is significant in the west and northeast region. The percent change in slope has higher increases over time. However, the changes are not statistically significant for the states with greater increase in slope (Texas, Oklahoma, and Georgia), whereas, for the other states, the slope change is significant. The last intervention, reopening, has an immediate increase in foot traffic, especially in the southeast region. This effect gradually decreases, and most states experience a negative change in foot traffic.

\FloatBarrier

\section*{Discussion}
Our findings suggest that the issuance of stimulus check caused a significant short-term increase in spending, but did not lead to much changes in foot traffic patterns. Over the long term, spending dropped back to the baseline level and foot traffic slowly returned to pre-lockdown level. This implies that stimulus check payment only had a temporary effect on spending, but it spurred a significant gradual increase in foot traffic. The impact of the second intervention, state-wide reopening policies, is less significant for spending, suggesting that lifting lockdown policies play limited roles in impacting spending. Even though events such as protests coincided with reopening in some states, they did not have significant effects on foot traffic. We see a flattening of the increasing trend for foot traffic in the data, which could be due to foot traffic approaching to the baseline level. This connects with the results from Cronin et al. \cite{Cronin2020} that a large fraction of social distancing is driven by voluntary precaution instead of legal restrictions, so lifting these restrictions does not have a huge impact on foot traffic.

Comparing the resulting maps of 50 states, we also see differences in the reaction to these policies in different regions in the US. The northeast region did not experience as much spending surge as the rest of the country, but it did have a large gradual increase over the next month. Although the effect of reopening is less significant on consumer spending, we find that the southeastern region tends to have a significant gradual increase in spending, whereas the other states have either negative change or insignificant positive change. We also see that the southeastern region has a greater and significant immediate increase in foot traffic after reopening, and these states have earlier reopening dates (before May 15th), which aligns with Alexander et al. \cite{alex} that early reopening states saw a stronger increase in foot traffic. The west coast, compared to the rest of the country, had smaller changes in spending and foot traffic responding the policies. 

The datasets we used have some limitations. The card transaction dataset is mainly from lower-income and younger consumers. Therefore, our findings regarding consumer spending primarily represent this subset of the US population. Since there was a significant immediate increase of spending after stimulus checks issuance, it connects with Coibion et al. \cite{coibion_2} that lower-income households were more likely to spend their stimulus checks, whereas higher-income individuals were more likely to save the money. The foot traffic data is from a population sample which is a panel of opt-in, anonymized smartphone devices, and is well balanced across US demographics and geographies. Because of the difference in population representation between the two datasets, we did not directly correlate between their trends.

Our results can be useful to policymakers when designing post-pandemic economic recovery plans. Providing financial support is a great way of boosting spending in the short run, but the impact is temporary. This contributes to a broader body of literature on the impact of fiscal policies during recessions, such as the Great Recession in 2008. For example, Fetai \cite{fetai} showed that fiscal policies can increase spending thus shortening the length of the financial crisis, and Reichling and Whalen \cite{Reichling} analyzed the direct and indirect effects of fiscal policies. While reopening businesses did not lead to a significant increase in spending and foot traffic, lockdowns took a heavy toll on many small businesses. More research is needed to analyze the longer-term effect of economic reopening on small businesses in light of the changing trends in spending and foot traffic.

Our findings can also inform businesses when they react to government policies during a pandemic. Since reopening plans may have a mild impact on foot traffic, businesses should still strive for connecting with customers online if applicable. Colleen et al. \cite{kirk} found that consumers' purchase decisions are strongly influenced by how brands react to the pandemic. Future work could examine how consumer risk perception affects their response (e.g., spending and foot traffic) to major events (e.g., implementations of policies and interventions) during the pandemic.

\section*{Acknowledgement}
We thank SafeGraph Inc. and Facteus for donating foot traffic and card transactions data for research purposes.

\bibliography{citations.bib}

\subsection*{Appendix}
For spending data, the percent of level change and slope change based on equation \ref{eq:1} are:
\\
Percent level change 1 $= \frac{\beta_2}{\beta_0 + \beta_1 \cdot time_t}$
\\
Percent level change 2 = 
$$\frac{\beta_4}{\beta_0 + \beta_1 \cdot time_t + [(\beta_1 + \beta_3) \cdot period_2]+ \beta_2}$$

Percent slope change 1 = $\frac{|\beta_3 |}{|\beta_1|} \cdot sign(\beta_3) $

Percent slope change 2 = $\frac{|\beta_5 |}{|\beta_1 + \beta_3|} \cdot sign(\beta_5)$,

\noindent where $period_2$ denotes the length of the period between the first and second intervention. 

For foot traffic data, the percent of level change and slope change based on equation \ref{eq:2} are:

Percent Level change 1 = $\frac{\beta_2}{\beta_0 + \beta_1 \cdot time_t}$

Percent level change 2 = 
$\frac{\beta_4}{\beta_0 + \beta_1 \cdot time_t + [(\beta_1 + \beta_3) \cdot period_2) + \beta_2]}$

Percent level change 3 = \\
$\frac{\beta_6}{\beta_0 + \beta_1 \cdot time_t + [(\beta_1 + \beta_3) \cdot period2 + \beta_2] + [(\beta_1 + \beta_3 + \beta_5) \cdot period_3] + \beta_4}$

Percent slope change 1 = $\frac{|\beta_3 |}{|\beta_1|} \cdot sign(\beta_3) $

Percent slope change 2 = $\frac{|\beta_5 |}{|\beta_1 + \beta_3|} \cdot sign(\beta_5)$

Percent slope change 3 = $\frac{|\beta_7 |}{|\beta_1 + \beta_3 + \beta_5|} \cdot sign(\beta_5)$,

\noindent where $period_2$ denotes the length of the period between the first and second intervention, and $period_3$ denotes the length of the period between the second and third intervention.

\end{document}